\documentclass[prd,onecolumn,nofootinbib,showpacs,showkeys]{revtex4}
\newcommand\gothfamily{\usefont{U}{ygoth}{m}{n}}
\DeclareTextFontCommand{\textgoth}{\gothfamily}

\begin{document}

\title{A VARIATIONAL FORMULATION OF RELATIVISTIC HYDRODYNAMICS}

\author{{\bf Nikodem J. Pop\l awski}}

\affiliation{Department of Physics, Indiana University, Swain Hall West, 727 East Third Street, Bloomington, IN 47405, USA}
\email{nipoplaw@indiana.edu}

\noindent
{\em Physics Letters A}\\
Vol. {\bf 373}, No. 31 (2009) pp. 2620--2621\\
\copyright\,Elsevier B. V.
\vspace{0.4in}

\begin{abstract}
We combine Taub's and Ray's variational approaches to relativistic hydrodynamics of perfect fluids into another simple formulation.
\end{abstract}


\maketitle

The action for a relativistic perfect fluid in a gravitational field is given by the integral \cite{Ray}
\begin{equation}
S=\frac{1}{c}\int d^4x\sqrt{-g}\biggl(-\frac{1}{2\kappa}R-\rho c^2-\rho\epsilon-\frac{1}{2}\mu_1(g_{\mu\nu}u^\mu u^\nu-1)+\mu_2\frac{d\sigma}{d\lambda}\biggr),
\label{act}
\end{equation}
where $R$ is the Ricci scalar, $g$ is the determinant of the metric tensor $g_{\mu\nu}$, $\rho$ is the rest density of the fluid (the number of particles per unit volume in the mean frame of reference of these particles), $\epsilon(\rho,\sigma)$ is the internal energy density per unit mass of the fluid, $u^\mu=\frac{dx^\mu}{d\lambda}$ is the four-velocity vector tangent to fluid-particle paths parametrized by $\lambda$, $\sigma$ is the entropy density of the fluid, and $\mu_1$ and $\mu_2$ are Lagrange multipliers that ensure the normalization of $u^\mu$ and the conservation of entropy for each particle, respectively.
The variation of the action (\ref{act}) is thus
\begin{eqnarray}
& & \delta S=\frac{1}{c}\int d^4x\sqrt{-g}\biggl(-\frac{1}{2\kappa}\Bigl(R_{\mu\nu}-\frac{1}{2}Rg_{\mu\nu}\Bigr)\delta g^{\mu\nu}-\mu_1u_\mu u^\nu(\delta x^\mu)_{;\nu}-\Bigl(c^2+\epsilon+\frac{p}{\rho}\Bigr)\delta\rho \nonumber \\
& & -\rho T\delta\sigma+\frac{1}{2}(\rho c^2+\rho\epsilon)g_{\mu\nu}\delta g^{\mu\nu}+\frac{1}{2}\mu_1u_\mu u_\nu\delta g^{\mu\nu}+\mu_2u^\mu(\delta\sigma)_{,\mu}\biggr),
\label{var}
\end{eqnarray}
where we use the first law of thermodynamics, $d\epsilon=Td\sigma+\frac{p}{\rho^2}d\rho$ ($T$ - temperature, $p$ - pressure), and the identity $\delta u^\mu=u^\nu (\delta x^\mu)_{;\nu}$.
The semicolon denotes covariant differentiation with respect to the Christoffel symbols and the comma denotes partial differentiation.
Integrating by parts in Eq. (\ref{var}) gives
\begin{eqnarray}
& & \delta S=\frac{1}{c}\int d^4x\sqrt{-g}\biggl(-\frac{1}{2\kappa}\Bigl(R_{\mu\nu}-\frac{1}{2}Rg_{\mu\nu}\Bigr)\delta g^{\mu\nu}+(\mu_1u_\mu u^\nu)_{;\nu}\delta x^\mu-\Bigl(c^2+\epsilon+\frac{p}{\rho}\Bigr)\delta\rho \nonumber \\
& & -\rho T\delta\sigma+\frac{1}{2}(\rho c^2+\rho\epsilon)g_{\mu\nu}\delta g^{\mu\nu}+\frac{1}{2}\mu_1u_\mu u_\nu\delta g^{\mu\nu}-(\mu_2u^\mu)_{;\mu}\delta\sigma\biggr).
\label{part}
\end{eqnarray}

In the integral (\ref{act}) we do not write explicitly, as in \cite{Ray}, a Lagrange-multiplier term that corresponds to the conservation of particle-number density:
\begin{equation}
(\rho u^\mu)_{;\mu}=0.
\label{cont}
\end{equation}
Instead, we impose this condition through a relation between $\delta\rho$, $\delta g^{\mu\nu}$ and $\delta x^\mu$ in Eq. (\ref{part}) \cite{Taub}.
The equation of continuity (\ref{cont}) gives $\delta\int d^4x(\sqrt{-g}\rho u^\mu)_{,\mu}=0$ or $\delta(\sqrt{-g}\rho u^\mu)=0$, from which we obtain
\begin{equation}
\delta\rho=\frac{1}{2}\rho g_{\mu\nu}\delta g^{\mu\nu}-\rho u_\mu u^\nu(\delta x^\mu)_{;\nu}.
\label{const}
\end{equation}
Substituting Eq. (\ref{const}) to (\ref{part}) gives
\begin{eqnarray}
& & \delta S=\frac{1}{c}\int d^4x\sqrt{-g}\biggl(-\frac{1}{2\kappa}\Bigl(R_{\mu\nu}-\frac{1}{2}Rg_{\mu\nu}\Bigr)\delta g^{\mu\nu}-\Bigl(\rho T+(\mu_2u^\mu)_{;\mu}\Bigr)\delta\sigma \nonumber \\
& & +\Bigl((\mu_1-\rho c^2-\rho\epsilon-p)u_\mu u^\nu\Bigr)_{;\nu}\delta x^\mu-\frac{1}{2}pg_{\mu\nu}\delta g^{\mu\nu}+\frac{1}{2}\mu_1u_\mu u_\nu\delta g^{\mu\nu}\biggr).
\label{integ}
\end{eqnarray}
Due to Hamilton's principle of least action, the action (\ref{integ}) vanishes for arbitrary variations $\delta\sigma$, $\delta x^\mu$ and $\delta g^{\mu\nu}$.
As a result we obtain, with the use of Eq. (\ref{cont}),
\begin{eqnarray}
& & T=-u^\nu\Bigl(\frac{\mu_2}{\rho}\Bigr)_{,\nu},
\label{temp} \\
& & u^\nu\biggl(\frac{\Delta}{\rho}u_\mu\biggr)_{;\nu}=0,
\label{Lagr} \\
& & R_{\mu\nu}-\frac{1}{2}Rg_{\mu\nu}=\kappa(\mu_1 u_\mu u_\nu-pg_{\mu\nu}),
\label{Ein}
\end{eqnarray}
where
\begin{equation}
\Delta=\mu_1-\rho c^2-\rho\epsilon-p.
\label{delta}
\end{equation}

The contracted Bianchi identity applied to the Einstein equations (\ref{Ein}) gives, together with Eq. (\ref{cont}), 
\begin{equation}
(\rho c^2+\rho\epsilon+p+\Delta)u^\nu u^\mu_{\phantom{\mu};\nu}=p_{,\nu}(g^{\mu\nu}-u^\mu u^\nu).
\label{iden}
\end{equation}
If $\Delta\neq 0$ then Eqs. (\ref{Lagr}) and (\ref{iden}) are in general not compatible because they constitute 8 constraints for only 5 independent quantities: 3 components of $u^\mu$, $\Delta$, and $\epsilon$ or $p$ (which are related by the equation of state of the fluid).
The compatibility of Eqs. (\ref{Lagr}) and (\ref{iden}) requires $\Delta=0$; Eq. (\ref{Lagr}) is thus satisfied identically and Eq. (\ref{iden}) gives 4 constraints for 4 independent quantities: 3 components of $u^\mu$ and $\epsilon$ (or $p$).\footnote{
For the equation of state of a dust, $p=0$, Eqs. (\ref{Lagr}) and (\ref{iden}) are compatible.
In this case, Eq. (\ref{iden}) reduces to the geodesic equation $u^\nu u^\mu_{\phantom{\mu};\nu}=0$, which, with Eq. (\ref{Lagr}), gives $u^\nu\bigl(\frac{\Delta}{\rho}\bigr)_{,\nu}=\frac{d}{d\lambda}\bigl(\frac{\Delta}{\rho}\bigr)=0$.
Consequently, the presented variational principle has a limitation for this case because $\Delta$ is not determined by the equation of motion; it is only constrained such that the ratio $\frac{\Delta}{\rho}$ is conserved for each fluid particle (including $\Delta=0$ as a solution).
}
Substituting the compatibility condition $\Delta=0$ to Eq. (\ref{iden}) gives relativistic Euler's equation of motion \cite{Ray,Taub}:
\begin{equation}
(\rho c^2+\rho\epsilon+p)u^\nu u^\mu_{\phantom{\mu};\nu}=p_{,\nu}(g^{\mu\nu}-u^\mu u^\nu),
\label{Eul}
\end{equation}
which, in the limit $c\rightarrow\infty$, reproduces the continuity and Euler's equation in nonrelativistic fluid mechanics \cite{LL6}.

Relativistic hydrodynamics of perfect fluids can be formulated from a variational principle with relativistic Clebsch's velocity potentials as dynamical variables \cite{pot}, but the presented formulation, like the methods of Taub and Ray, requires a smaller number of independent field components.
Taub's formulation uses the Helmholtz free energy density in the action and imposes the conservation of particle-number density through a relation between the variations of $\delta\rho$, $\delta g^{\mu\nu}$ and $\delta x^\mu$ \cite{Taub,SS}, while Ray's formulation uses the internal energy density in the action and imposes the conservation of particle-number density through a Lagrange-multiplier term \cite{Ray,Elze}.
There also exists a Hamiltonian formulation of relativistic perfect fluids based on scalar fields, which is free of any constraints \cite{Ham}.

The difference between the presented formulation and the variational approaches of Taub and Ray is that we use in the action (\ref{act}) the internal energy density $\epsilon$, as in \cite{Ray}, and use the variational relation (\ref{const}) to impose the conservation of particle-number density, as in \cite{Taub}.
Because of this relation, we do not need to introduce Lin's constraint \cite{Ray,Lin}, without which isentropic flows described by the action for a perfect fluid with a particle-conservation Lagrange multiplier are irrotational \cite{Herivel}.
We also, as in Ray's formulation, impose the conservation of entropy through a Lagrange-multiplier term in the action.
In Taub's formulation, this conservation is derived from the variation of the quantity $\alpha$ related to $T$ via $\delta T=\frac{\partial\delta\alpha}{\partial\lambda}$ \cite{Taub}.
We vary the particle paths, $\delta x^\mu$ (Lagrangian picture) \cite{Taub}, and not the four-velocity, $\delta u^\mu$ (Eulerian picture) \cite{Ray,Chiueh}, although both approaches are equivalent in the presented formulation because we do not regard $\delta x^\mu$ as the coordinate variation and thus $g_{\mu\nu}$ does not change under $\delta x^\mu$.
The metric tensor changes in \cite{Taub} under $\delta x^\mu$ and the corresponding field equation is the conservation of the dynamical energy-momentum tensor \cite{LL2}.

Our variational formulation of relativistic hydrodynamics of perfect fluids can be generalized to the Einstein-Cartan spacetime \cite{RS}, and to the general metric-affine spacetime \cite{Smal}.
In the presence of a general affine connection, the continuity relation for a fluid takes its general-relativistic form (with the Christoffel symbols in the covariant derivative) (\ref{cont}) \cite{Smal}, otherwise the field equation resulting from the variation of the connection would impose unphysical constraints on the four-velocity \cite{Hehl}.
This field equation brings the connection to the Levi-Civita form combined with the torsion trace which does not appear in the symmetric part of the Ricci tensor so the Einstein equations remain unchanged.
Therefore the presented formulation is valid also in the spacetime with a general affine connection, that is, with torsion and nonmetricity \cite{tor}.
It should not be difficult to generalize this formulation to hyperfluids, which are continuous media that carry momentum and hypermomentum \cite{OT}.

\end{document}